\documentclass[12pt]{iopart}

\usepackage{iopams}  
\usepackage{graphicx}
\usepackage{subfigure}
\usepackage{cite}
\begin{document}

\title[Article Title]{Microscopic Insights for Beyond Room-Temperature Ferromagnetism in Ni doped Two-Dimensional Fe$_5$GeTe$_2$}

\author{Sukanya Ghosh, Soheil Ershadrad and Biplab Sanyal*}

\address{Department of Physics and Astronomy, Uppsala University, Box-516, 75120, Uppsala, Sweden}
\ead{*biplab.sanyal@physics.uu.se}
\vspace{10pt}

\begin{abstract}
Enhancement of Curie temperature ($T_\mathrm{C}$) of two-dimensional (2D) magnets is immensely desirable for room temperature spintronic applications. Fe$_{5}$GeTe$_{2}$ is an exceptional van der Waals metallic ferromagnet due to its tunable physical properties and relatively higher $T_\mathrm{C}$ than other 2D magnets. Using density functional theory combined with dynamical mean field theory and Monte Carlo simulations, we show that the $T_\mathrm{C}$ of Fe$_{5}$GeTe$_{2}$ monolayer can increase well-above room temperature by substitutional doping with Ni. It is found that two specific sublattices (Fe1 and Fe4) are the first and second most energetically preferred  occupation sites for Ni. $T_\mathrm{C}$ of Fe$_{5-\delta}$Ni$_{\delta}$GeTe$_{2}$ increases up to $\sim$ 400 K at $\delta \sim$20\%. Exchange interactions between particular Fe5-Fe4 pairs play a dominating role in tuning the transition temperature, influenced by doping-induced structural distortions. Finally, we highlight the effect of dynamical electron correlation in site-specific electronic structure and quasi-particle mass of Fe-$d$ orbitals with varying Ni doping.

\end{abstract}




\maketitle









\section{Introduction}\label{sec1}

Atomically thin, layered quasi-two-dimensional (2D) van der Waals (vdW) crystals exhibit exceptional physical properties\cite{Gibertini_2019,Cheng_Science_2019}. 
Regarding 2D magnets, however, according to the Mermin-Wagner theorem, an intrinsic long-range magnetic order can not exist in the isotropic 2D limit because strong thermal fluctuations prohibit continuous symmetries to break spontaneously\cite{Mermin_1966}. The presence of weak magnetic anisotropy is sufficient to open up a sizable gap in the magnon spectra, causing long-range magnetic order to persist in materials with dimension $D\le2$ at a finite temperature. 
Among the newly discovered vdW magnetic materials\cite{Gibertini_2019,Burch_Nature_2018,Zollner_PRL_2020,CGT_Nat_Commun,CGT_Nat_Elect,Lee_2020_CrI3,Wang_PRL_CrI3,Jiang_Nature,SG_Nanoscale,GHOSH2019}, metallic  Fe$_{n}$GeTe$_{2}$($n=3, 4, 5)$\cite{SE_jpcl,SG_FGT2022,Sci_adv_FGT_family,Ramesh_PRL_2022,Bing2023} systems, commonly referred to as FGT are quite special. Ferromagnetism close to room temperature increases their demand for spintronic applications. This can be further enhanced by applying pressure\cite{CGT_PRL_2021,FGT_strain_2020}, gating\cite{FGT_gate_2018,Kim_2021}, carrier doping\cite{FGT_NL_2019,may2020Co}, ion intercalation\cite{CGT_JACS_2019}, etc. It has also been found that substitutional doping of Fe$_{5}$GeTe$_{2}$ with cobalt increases the magnetic ordering temperature to $\sim$ 360 K influencing the magnetic ground state, interlayer stacking and magnetic textures\cite{may2020Co,Ramesh_PRM_2022,Ramesh_SciAdv_2022,Co_APL}. A recent study by Chen et al. has reported an enhancement of ferromagnetism in bulk Fe$_{5-\delta}$GeTe$_{2}$ up to 478 K with Ni doping\cite{Ramesh_PRL_2022}. The authors showed that Ni doping triggers structural modifications, influences saturation magnetization, and affects $T_\mathrm{C}$. However, a detailed description of microscopic mechanisms determining the observed trend for $T_\mathrm{C}$ in Fe$_{5-\delta}$Ni$_{\delta}$GeTe$_{2}$ is still lacking. Especially, the exact nature of magnetic interactions governing $T_\mathrm{C}$ should be unveiled. How the tuning of exchange interactions caused by structural modifications in Fe$_{5-\delta}$Ni$_{\delta}$GeTe$_{2}$ determines $T_\mathrm{C}$, demands a thorough investigation.

 In this Letter, performing first-principles calculations, we investigate the rationals responsible for the enhancement of $T_\mathrm{C}$ well above room temperature by substitutional doping with Ni in Fe$_{5}$GeTe$_{2}$ monolayer. We find out which particular Fe sites are more prone to host Ni dopant. Our study explains how the tuning of exchange interactions between certain Fe pairs, caused by structural modifications, plays the dominating role to increase $T_\mathrm{C}$ up to a certain doping. Reduction in $T_\mathrm{C}$ for higher doping is caused by the replacement of magnetic Fe atom by nonmagnetic Ni, however, the dominating exchange still remains ferromagnetic, as observed in experiment\cite{Ramesh_PRL_2022}, but in contrast with a recent DFT study\cite{Hu_PRB}.

 A $\sqrt 3 \times \sqrt 3$ cell of Fe$_{5}$GeTe$_{2}$ monolayer in UDU (Up-Down-Up) configuration of Fe atoms is considered in this study, where two (Fe1U) and one (Fe1D) Fe atoms are situated directly above and below Ge, respectively\cite{SE_jpcl,Ramesh_PRB_2020,may2019FM,may2019physical}. We investigate magnetic and electronic properties of the energetically favored configurations, determined by comparing the total energies of Fe$_{5-\delta}$Ni$_{\delta}$GeTe$_{2}$ monolayers varying the position(s) of Ni dopant(s). 

\begin{figure*} [ht]
\begin{center}
\includegraphics[width=0.85\textwidth] {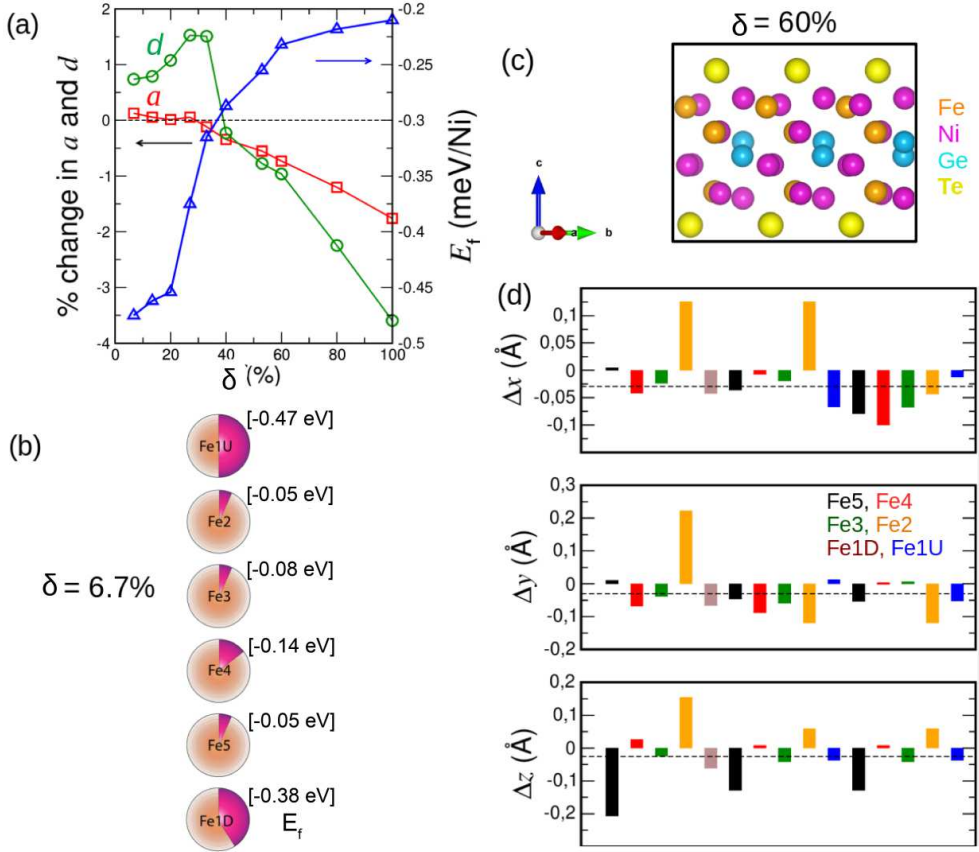}
\caption{ (a) Percentage change of in-plane lattice parameter $a$ (red squares), thickness $d$ (green circles) and formation energy $E_\mathrm{f}$ (blue triangles), respectively, for Fe$_{5-\delta}$Ni$_{\delta}$GeTe$_{2}$ monolayer with Ni doping ($\delta$). (b) Schematics showing formation energy ($E_\mathrm{f}$) of Fe$_{5-X}$Ni$_{X}$GeTe$_{2}$ for different Ni occupation site. Purple area indicates the tendency of a Fe sublattice to get substituted with Ni, numbers show $E_\mathrm{f}$ for different Fe sites at $\delta=$6.7\%. (c) Side view of Fe$_{5-\delta}$Ni$_{\delta}$GeTe$_{2}$ at $\delta=60\%$. The height of histograms shows rumpling of each Fe/Ni atom along $x$, $y$ and $z$ directions present at the unit cell of $\delta=60\%$ wrt the undoped monolayer, horizontal dashed line shows the average rumpling. }
\label{Fig1}
\end{center}
\end{figure*}

  Our results show that the in-plane lattice parameter $a$ remains almost unaltered compared to the undoped system till $\delta=27\%$, and starts to reduce when $\delta>30\%$, see Fig.~\ref{Fig1}(a). However, the thickness $d$ of the monolayer increases with doping up to $\delta=20\%$ but reduces for $\delta\ge 40\%$ as shown in Fig.~\ref{Fig1}(a). Substitutional doping of Ni in Fe$_{5}$GeTe$_{2}$ monolayer becomes energetically less favored with increase in $\delta$. This is evident in Fig.~\ref{Fig1}(a) (blue triangles) where the formation energy $E_\mathrm{f}$ per Ni dopant is observed to increase with $\delta$. It is also worth noting that during the substitutional doping, the replacement of Fe1 species with Ni is energetically more favored than other Fe sites. 
  Between $\delta=$ 6.7\%(Fe1U) and 20\%(Fe1U+Fe1U+Fe1D) only Fe1 sublattice gets substituted by Ni.  After Fe1, the next energetically favored occupation site for Ni dopant is Fe4. There is a significant increase in $E_\mathrm{f}$ between $\delta=$20\% and 27\%, when one of the Fe4 atoms gets substituted together with Fe1U and Fe1D species. The presence of Ni causes an excess of electrons (Fig~S1), which might cause the lowering of $E_\mathrm{f}$. For $\delta\ge$ 33\%, Fe atoms belonging to other Fe sublattices (Fe2, Fe5 and Fe3) start to get substituted along with Fe1 and Fe4. Fig.~\ref{Fig1}(b) shows how $E_\mathrm{f}$ varies when Ni  substitutes different Fe sublattices at $\delta=6.7\%$. 

Structural distortion or rumpling (along $x$, $y$ and $z$ directions) in the monolayer increases with Ni-doping (Fig.~S1). Fig.~\ref{Fig1}(c) shows the side view when $\delta=$ 60\%. The height of the histograms in Fig.~\ref{Fig1}(d) shows the difference in $x$, $y$ and $z$ coordinates between Fe or Ni atoms present in 60\% Ni-doped and undoped systems for each (Ni/Fe) site of $\sqrt3\times\sqrt3$ cell. $\delta=60\%$ causes significant rumpling of the atoms present at the sites of Fe5 (along $z$) and Fe2 sublattices. Negative value of average rumpling (dashed horizontal line) supports the compression of cell parameters (both $a$ and $d$) with Ni doping, as we see in Fig.~\ref{Fig1}(a). 
It should be noted that Ref.~\citenum{Ramesh_PRL_2022} also finds a reduction in layer thickness with an increase in Ni doping. Apart from substitutional doping, Ni can occupy any vacant site of $\sqrt{3}\times\sqrt{3}$ cell of Fe$_{5-\delta}$GeTe$_{2}$, including vdW gap between different layers, such scenario can be present in experiments performed at a finite temperature\cite{Ramesh_PRL_2022}.  

\begin{figure} [b]
\begin{center}
\includegraphics[width=0.65\textwidth] {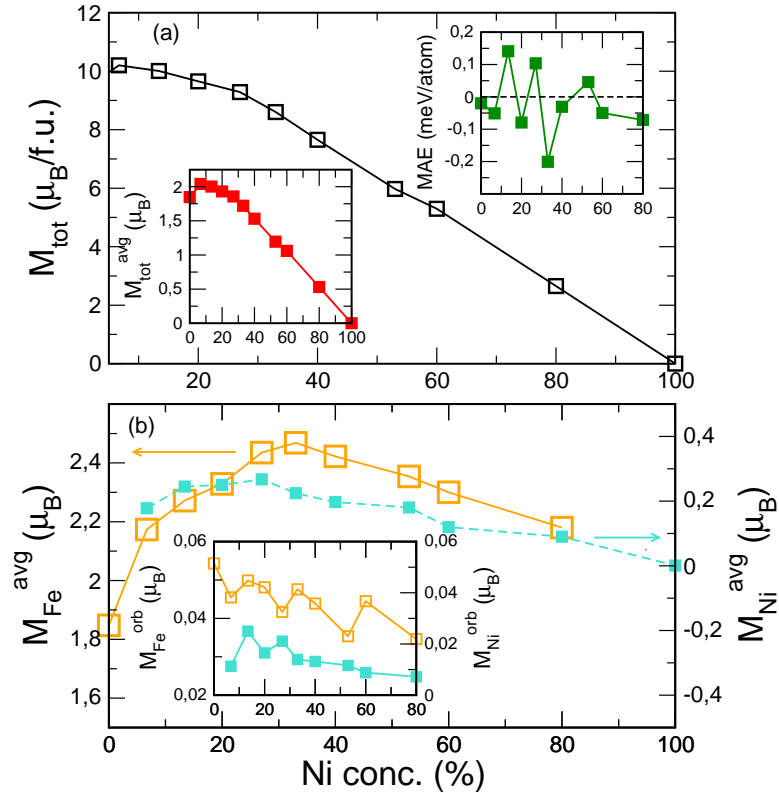}
\caption{(a) Total magnetic moment (M$_\mathrm{tot}$) of Fe$_{5-\delta}$Ni$_{\delta}$GeTe$_{2}$ monolayer as a function of doping concentration $\delta$. Insets with red and green squares show the average of total magnetic moment $M_{tot}^{avg}$ and variation of magnetic anisotropy energy (MAE) as a function of $\delta$, respectively. (b) Variation of average magnetic moment of Fe (M$_\mathrm{Fe}^\mathrm{avg}$, orange squares) and  Ni (M$_\mathrm{Ni}^\mathrm{avg}$, cyan squares) atoms plotted  with $\delta$. Inset shows average orbital moment $M_\mathrm{orb}^\mathrm{avg}$ for Fe (orange) and Ni (cyan) for different $\delta$.  }
\label{Fig2}
\end{center}
\end{figure}

\begin{figure} [ht]
\begin{center}
\includegraphics[width=0.55\textwidth] {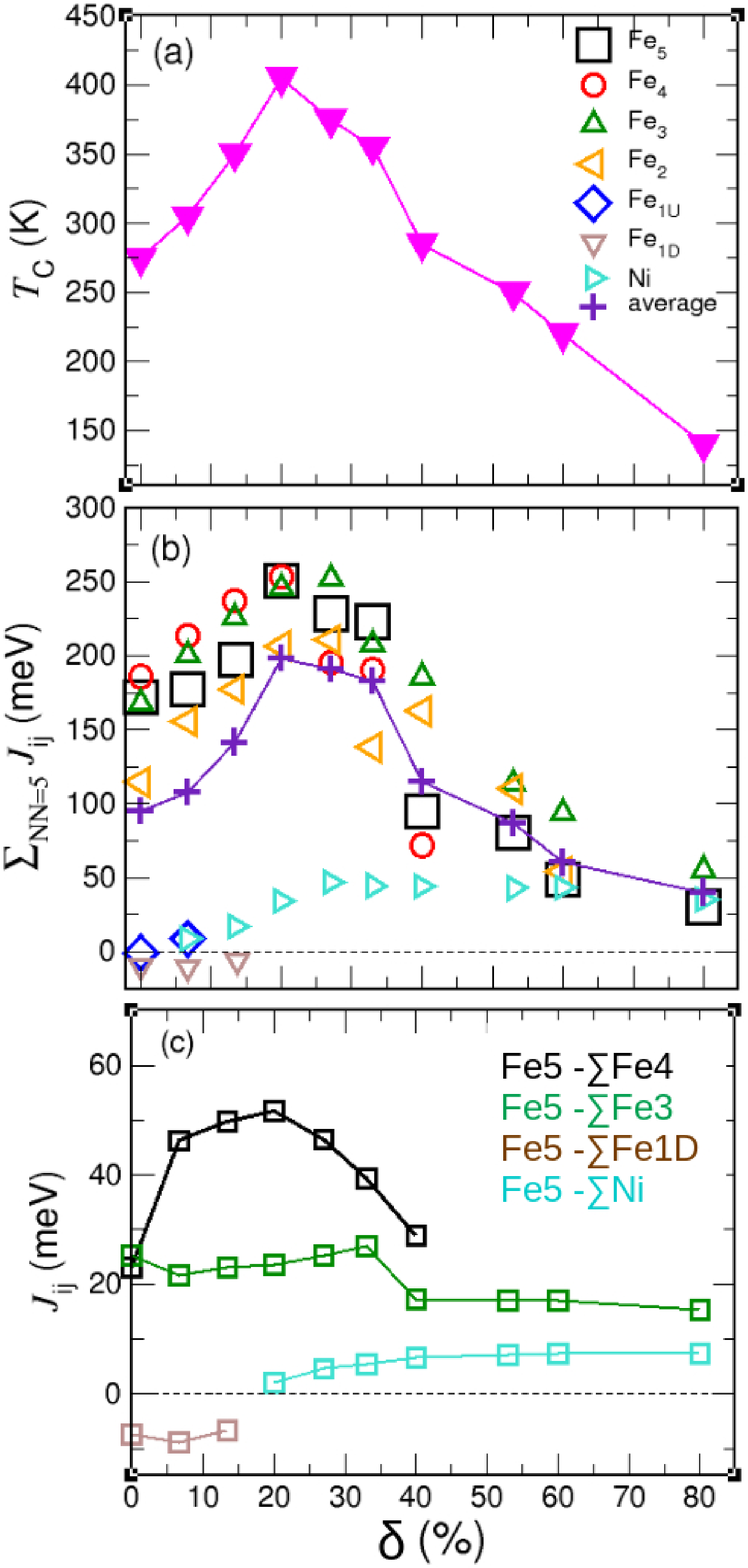}
\caption{(a) Variation of $\sum_{NN=5}{J_{ij}}$ with $\delta$ for each Fe sublattice. $\sum_{NN=5}{J_{ij}}$ is sum of isotropic symmetric exchange interactions $J_{ij}$ over the first five nearest neighbors. (b) Variation of $T_\mathrm{C}$ with $\delta$. (c) $J_{ij}$ interactions between $i=$Fe5 and $j=\sum$Fe4, $\sum$Fe3, $\sum$Fe1D and $\sum$Ni with $\delta$. 
}
\label{Fig3}
\end{center}
\end{figure}

 In our previous study, we showed that the inclusion 3of dynamical electron correlation effect is an appropriate approach to determine the magnetic moment, exchange interactions and $T_\mathrm{C}$ of the Fe$_{n}$GeTe$_{2}$ systems, compared to the standard GGA and GGA+U methods\cite{SG_FGT2022}. Therefore, we perform charge self-consistent dynamical mean-field theory (DFT+DMFT) calculations as implemented in the FP-LMTO code RSPt \cite{Oscar,Wills2000} to investigate the magnetic and electronic properties of Fe$_{5-\delta}$Ni$_{\delta}$GeTe$_{2}$ monolayer. The computational details are mentioned in SI. 
 
 Doping with Ni influences the magnetism of Fe$_{5-\delta}$Ni$_{\delta}$GeTe$_{2}$ monolayer. The total spin moment $M_\mathrm{tot}$ reduces as a function of doping ($\delta$). Fig.~\ref{Fig2}(a) shows that $M_\mathrm{tot}$ decreases from 10 $\mu_{B}$ to 0 $\mu_{B}$ from $\delta=0\%$ to 100\%. These results are in good agreement with the saturation magnetic moment of bulk Fe$_{5-\delta}$Ni$_{\delta}$GeTe$_{2}$. \cite{Ramesh_PRB_2020} 
It is interesting to note that the direction of easy axis or magnetic anisotropy energy, MAE (= $E_{||}-E_{\perp}$) oscillates between in-plane and out-of-plane directions with $\delta$. However, for most of the doped systems, the easy axis lies in the $xy$-plane, see green squares presented as an inset in Fig.~\ref{Fig2}(a), which is in agreement with experiment\cite{Ramesh_PRL_2022}. The strength of MAE for Fe$_{5-\delta}$Ni$_{\delta}$GeTe$_{2}$ monolayers is much weaker than pristine Fe$_{3}$GeTe$_{2}$ and Fe$_{4}$GeTe$_{2}$ monolayers\cite{SG_FGT2022,Kim_2021}.  The trend observed for MAE at lower $\delta$ values can be correlated with the value of orbital moments obtained for different directions of spin axis\cite{Bruno_1989}, see Table~S2. It is worth noting that the switching of the easy axis is observed in Fe$_{5-\delta}$GeTe$_{2}$ with Co doping and electrical gating\cite{may2020Co,F5GT_Nat_2022}. Experiments report switching of easy axis for bulk Fe$_{5}$GeTe$_{2}$ depending on the Fe concentration\cite{may2019ferromagnetism,Ramesh_PRB_2020}. The average spin moment of Fe atoms first increases with doping, becomes maximum at $\delta=33\%$, then reduces, see Fig.~\ref{Fig2}(b). Similar to Ref.~\citenum{Ramesh_PRL_2022}, our calculations also find that Ni dopants carry negligible spin moment (cyan squares in Fig.~\ref{Fig2}(b)) and are not responsible for the origin or tuning of ferromagnetism. The average orbital moment $M_\mathrm{orb}^\mathrm{avg}$ of Fe remains between 0.05 and 0.04 till $\delta$=80\%, see inset of Fig.~\ref{Fig2}(b). The orbital moment of Ni falls in the range of 0.01-0.02 and reduces for higher $\delta$.

\begin{figure} [t]
\begin{center}
\includegraphics[width=0.8\textwidth] {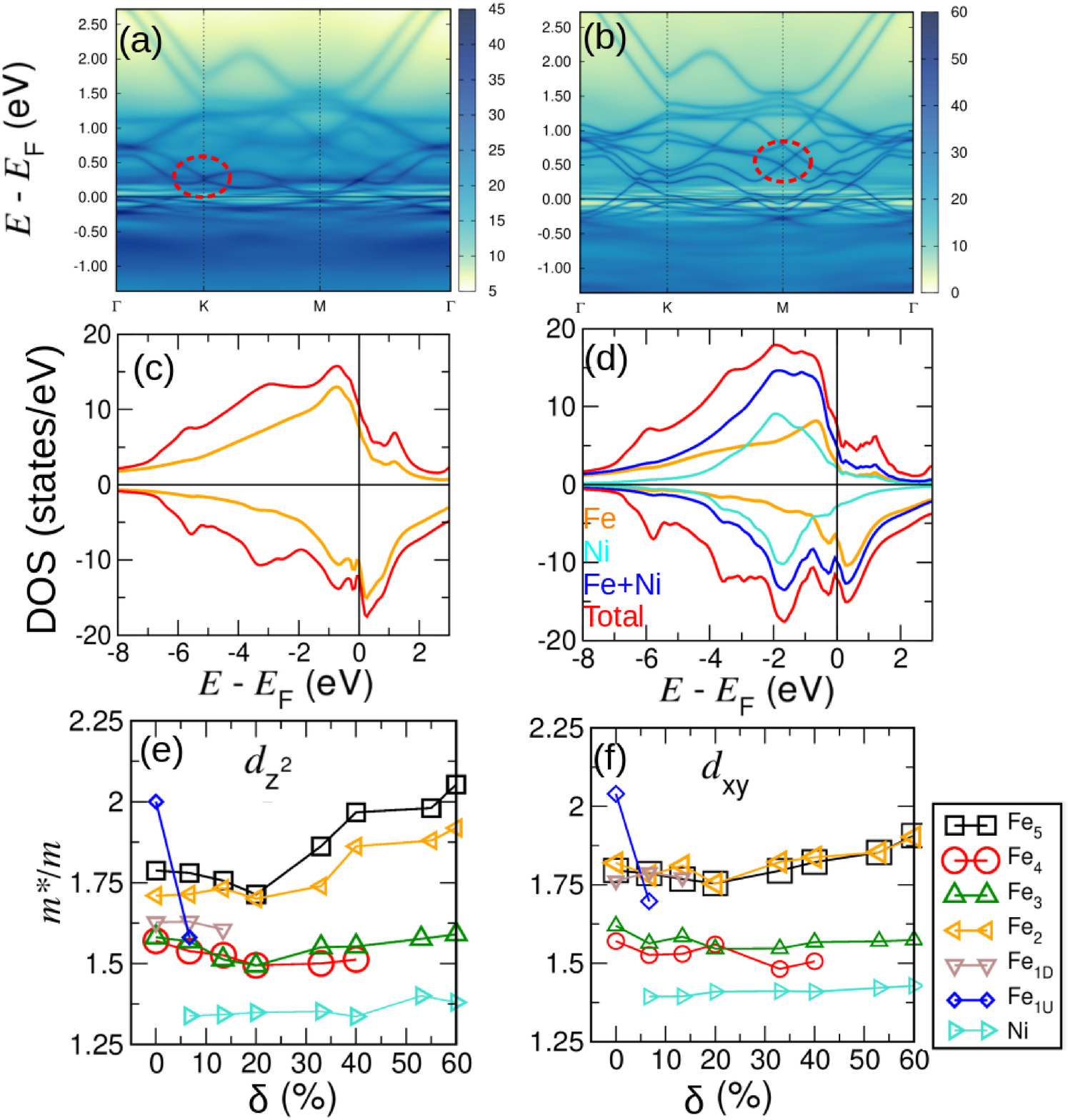}
\caption{Spectral function plots for $\delta=$ (a)0\% and (b) 40\%. The Dirac-cone-type features present at the high-symmetry points K (a) and M (b) are highlighted by the red circles. (c) and (d) DOS at $\delta=0\%$ and 40\%, respectively. Effective mass, $m*/m$ for (e) Fe-$d_{z}^{2}$ and (f) $d_{xy}$ for different Fe sublattices and Ni with $\delta$. These results are obtained at $T=155$ K.}
\label{Fig4}
\end{center}
\end{figure}    

We calculate the isotropic symmetric ($J_{ij}$) and antisymmetric ($D_{ij}$) exchange interactions present in Fe$_{5-\delta}$Ni$_{\delta}$GeTe$_{2}$ monolayer varying $\delta$. Incorporating $J_{ij}$, $D_{ij}$ and MAE in the Heisenberg Spin-Hamiltonian (Eq.~S4), $T_\mathrm{C}$ is computed performing Monte Carlo simulations. Fig.~\ref{Fig3}(a) shows there is a monotonic increase of $T_\mathrm{C}$ up to $\delta=20\%$, then it reduces. Qualitative trend of $T_\mathrm{C}$ vs. $\delta$ plotted in Fig.~\ref{Fig3} agrees well with experimental reports on bulk Fe$_{5-\delta}$Ni$_{\delta}$GeTe$_{2}$\cite{Ramesh_PRL_2022}. The critical $\delta$ value at which $T_\mathrm{C}$ of the monolayer becomes maximum, is not exactly the same as observed in the experiment. This slight discrepancy happens because in experiments, during the doping process, Ni can be placed at any vacant position present in the bulk without replacing Fe, enhancing ferromagnetism and hence $T_\mathrm{C}$. However, the rational behind such trend of $T_\mathrm{C}$ in Fe$_{5-\delta}$Ni$_{\delta}$GeTe$_{2}$ is not addressed in the previous study. 

Comparing the strength of different magnetic interactions we expect $J_{ij}$ couplings must play the dominating role to determine $T_\mathrm{C}$, in agreement with our study on pristine FGT systems\cite{SG_FGT2022}. In order to investigate the tuning of $J_{ij}$ with $\delta$, we plot the $J_{ij}$ values summed over the first five nearest neighbors (NNs) for different $i$th Fe sublattices, see Fig.~\ref{Fig3}(b). We consider NN up to 5, because the $J_{ij}$ interactions decay significantly beyond that, see Figs.S4-S7. Most of the Fe sublattices show dominating ferromagnetic (FM) $J_{ij}$ interactions while Fe1D and Fe1U show antiferromagnetic (AFM) interactions, till $\delta=$13.4\%. 

     
 The $\sum_{NN=5}{J_{ij}}$ term for Fe5, Fe4, Fe3 and Fe2 first increases with $\delta$, becomes maximum for $\delta=$20\%, then reduces for higher concentration. $\sum_{NN=5}{J_{ij}}$ is plotted for each Fe species present in the $\sqrt{3}\times\sqrt{3}$ cell. A monotonic increase of $\sum_{NN=5}{J_{ij}}$ for each Fe sublattice (except Fe1U and Fe1D) is observed up to $\delta=20\%$. This happens because till $\delta=20\%$, except Fe1, the number of Fe sublattices present in the unit cell is 3. For $\delta>$20\%  the Fe atoms belonging to Fe4 sublattice start to get substituted in addition with Fe1U and Fe1D, as we see in Fig.~\ref{Fig1}(d). The replacement of magnetic Fe causes sharp reduction in $\sum_{NN=5}{J_{ij}}$ of Fe4 for $\delta > 20\%$. After 40\% all three Fe4 atoms get substituted with Ni, see red circles in Fig.~\ref{Fig3}(a). The gradual replacement of different Fe sublattices with Ni causes rapid lowering in $\sum_{NN=5}{J_{ij}}$ for $\delta \ge$ 33\%. In addition to possessing negligible magnetic moment (Fig.~\ref{Fig2}), Ni dopants have negligible contribution to the $J_{ij}$ interactions. The magnitude of $\sum_{NN=5}{J_{ij}}$ for Ni is $\sim$ 10 times smaller than the Fe sublattices for $\delta \le 20\%$. As the number of Ni atoms present in Fe$_{5-\delta}$Ni$_{\delta}$GeTe$_{2}$ increases with $\delta$, $\sum_{NN=5}{J_{ij}}$ for Ni becomes comparable with Fe for $\delta \ge 60\%$. For $i$=Ni, non-zero exchange couplings exist when $j$=Fe, otherwise, interactions between Ni themselves is rather weak. The violet symbols show the average variation of $\sum_{NN=5}{J_{ij}}$ with Ni doping, which increases from 0\% to 20\% and then reduces. The same investigation has been made for $D_{ij}$ interactions as well. For a given $\delta$, the magnitude of $\sum_{NN=5}{D_{ij}}$ for any $i$th Fe species is $\sim 10$ times smaller than $\sum_{NN=5}{J_{ij}}$ (Fig~S9).

     Further analysis of $J_{ij}$ couplings among individual Fe pairs reveals that the interactions between Fe5 and Fe4 play the dominating role. Fig~\ref{Fig3}(c) shows $J_{ij}$ interactions when $i=$ Fe5 and $j$th species are  considered to be the nearest neighbors of Fe5, i.e., Fe4, Fe3, Fe1D and Ni. We find, among these neighbors, exchange coupling between Fe5-$\sum$Fe4 plays dominating role to control or tune the $T_\mathrm{C}$. The $J_{ij}$ interaction between Fe5 and Fe4 increases with $\delta$ and then reduces for $\delta \ge 27\%$. The increase of $J_{54}$ up to $\delta=20\%$ occurs due to the following reasons: i) reduction of Fe5-Fe4 bond length with $\delta$ (Fig.~S2), ii) among the first five NNs of Fe5, three of them are Fe4 species with the strongest FM coupling. Therefore, $J_{54}=(J_{54_{1}}+ J_{54_{2}}+ J_{54_{3}})$ for $\delta\le20\%$. The reduction in $J_{54}$ for $\delta > 20\%$ occurs due to the gradual replacement of Fe4 with Ni, causing a decrease in the number of Fe4 belonging to the first five NN of Fe5, see Fig.~S3 for details. 
  No particular trend is observed for $J_{53}$ (green squares), up to $\delta=60\%$. $J_{ij}$ exchange interactions between Fe5 and Fe1D is AFM (brown squares), and replacement of Fe1D with Ni triggers FM exchange coupling (cyan squares). Comparing different $J_{5j}$ interactions we see $J_{54}$ follows the  similar trend as $\sum_{NN=5}{J_{5j}}$ (black squares) and average of $\sum_{NN=5}{J_{ij}}$ (purple symbols) in Fig.~\ref{Fig3}(b). Therefore, $J_{54}$ has major influence on $T_\mathrm{C}$, especially for $\delta \le 20\%$. From Fig.~\ref{Fig3}(a) - (c) we can establish the fact that $T_\mathrm{C}$ not only depends on the strength of $J_{ij}$ but also on the nearest neighbor (NN) distance or effective coordination number. 

  

Now we discuss how Ni doping modifies the electronic structure.
Figs.~\ref{Fig4}(a) and (b) show the spectral function $A(k,\omega)$ for $\delta=$0\% and 40\%. The main differences observed in $A(k,\omega)$ at these two values of $\delta$ are:i) at the high-symmetry point $K$, there is presence of Dirac-cone-type feature for $\delta$=0\%, which disappears at $\delta=40\%$, ii) at $M$, Dirac-cone is observed for $\delta=40\%$ which is not present for the undoped monolayer. The density of states (DOS) plots for $\delta=0\%$ and 40\% are plotted in Figs.~\ref{Fig4}(c) and (d), respectively, (see Fig.~S16 for details). The intensity of DOS projected on Fe reduces with Ni doping. The Fe states have dominating contribution close to $E_\mathrm{F}$, maximum intensity of Ni states arises away from $E_\mathrm{F}$. Similar to other FGT systems, the admixture of localized and itinerant electrons exists in Fe$_{5}$GeTe$_{2}$ as well\cite{Kondo_Nanolett,SG_FGT2022}. The effective mass ($m^{*}/m)_{l\sigma}$ provides the quantitative measurements of electronic correlation\cite{mass_1992} (see Eq.~S7). Both qualitative and quantitative trends of $m^{*}/m$ with $\delta$ remain almost unaltered for different Fe-$d$ states, see Figs.~\ref{Fig4}(e)-(f) and Fig.S17. $m^{*}/m$ of Fe reduces between $\delta=0\%$ and 20\%, then increases for $\delta>20\%$, especially for Fe5 and Fe2. ($m^{*}/m)_{l\sigma}$
of Fe5, Fe2 are larger than for Fe3, Fe4. This implies the $d$ electrons belong to Fe5 and Fe2 sublattices are more correlated than Fe3 and Fe4. Similar to Fe$_{5-\delta}$GeTe$_{2}$, site-dependence of Fe sublattices is present in Fe$_{3}$GeTe$_{2}$ as well\cite{Hund_F3GT}. Interestingly, the effective mass of Ni-$d$ is lower than Fe-$d$, signifying the Ni-$d$ states are less correlated or more delocalized than Fe-$d$. 


 In summary, we investigate the effect of Ni doping on the structural, electronic, and magnetic properties of $\sqrt{3}\times\sqrt{3}$ Fe$_{5-\delta}$Ni$_{\delta}$GeTe$_{2}$ monolayer using DFT+DMFT and Monte Carlo simulations. Our results show that $T_\mathrm{C}$ of the monolayer increases up to $\sim$400 K by substitutional doping with Ni. The variation in $T_\mathrm{C}$ of Fe$_{5-\delta}$Ni$_{\delta}$GeTe$_{2}$ monolayer with doping is in good agreement with a recent experimental report on bulk Fe$_{5-\delta}$Ni$_{\delta}$GeTe$_{2}$. Moreover, we identify the dominant exchange interactions responsible for the observed trend in $T_\mathrm{C}$. Our results show the structural modifications caused by the Ni dopant, thereby modifying the nearest neighbor distances and effective coordination numbers, which affect the dominating exchange couplings. Our study also shows how Ni-doping influences site-dependent spectral features and effective masses arising from electron correlation.
\\
\\

\textbf{Acknowledgments}

B.S. and S.G. acknowledge a postdoctoral grant from Carl Tryggers Stiftelse (CTS 20:378). B. S. acknowledges financial support from Swedish Research Council (grant no. 2022-04309). The computations were enabled in project SNIC 2022/3-30 by resources provided by the Swedish National Infrastructure for Computing (SNIC) at NSC, PDC, and HPC2N partially funded by the Swedish Research Council (Grant No. 2018-05973). B.S. acknowledges allocation of supercomputing hours by PRACE DECI-17 project `Q2Dtopomat' in Eagle supercompter in Poland and EuroHPC resources in Karolina supercomputer in Czech Republic and LUMI supercomputer in Finland. 

\section{References}
\bibliography{apssamp}


\end{document}